\documentclass[12pt, a4paper]{iopart}

\usepackage{dsfont}
\usepackage{graphicx}
\usepackage{subfigure}
\usepackage[sort&compress]{natbib}
\bibpunct{(}{)}{,}{authoryear}{}{,}
\usepackage{hyperref}

\newcommand{\argmaxAA}{\arg\!\max} % AlfC

\begin{document}

\title[Decoding of spinal motor neuron spike trains for estimating hand kinematics]{Unsupervised decoding of spinal motor neuron spike trains for estimating hand kinematics following targeted muscle reinnervation}

\author{Arash~Andalib$^1$, Dario~Farina$^2$, Ivan~Vujaklija$^2$, Francesco~Negro$^3$, Oskar~C~Aszmann$^4$, Rizwan~Bashirullah$^1$ and Jose~C~Principe$^1$}

\address{$^1$ Department of Electrical and Computer Engineering, University of Florida, Gainesville, FL 32611 USA}
\address{$^2$ Department of Bioengineering, Imperial College London, London, UK}
\address{$^3$ Department of Clinical and Experimental Sciences, Universit\`{a} degli Studi di Brescia, Italy}
\address{$^4$ Laboratory of Bionic Extremity Reconstruction, Division of Plastic and Reconstructive Surgery, University of Vienna, AUSTRIA}

\ead{andalib@ufl.edu}
\vspace{10pt}
\begin{indented}
\item[]December 2018
\end{indented}

\begin{abstract}
The performance of upper-limb prostheses is currently limited by the relatively poor functionality of unintuitive control schemes. This paper proposes to extract, from multichannel electromyographic signals (EMG), motor neuron spike trains and project them into lower dimensional continuous signals, which are used as multichannel proportional inputs to control the prosthetic's actuators. These control signals are an estimation of the common synaptic input that the motor neurons receive. We use the simplest of metric learning approaches known as principal component analysis (PCA), as a linear unsupervised metric projection to extract the spectral information of high dimensional data into the subspace of the prosthetic hand degrees of freedom. We also investigate the importance of a rotation in the projection space that best aligns the PCA subspace with the space of degrees of freedom of the prosthetic hand, to attempt to approximate the sparseness properties of the motor axes (which are orthogonal), while no sparseness is enforced by PCA. Proof of concept for the feedforward path (open loop) is given by successful estimation of concurrent movements with up to three degrees of freedom. We also analyze and quantify the performance of the proposed decoding algorithm under implementation constraints and hardware limitations and propose a space-time subsampling strategy, to maximize projection fidelity, in each extracted source over time. The results confirm that the proposed decoding approach can directly project the motor neuron spike trains into kinematics for simultaneous and proportional prosthesis control that can be extended to multiple degrees of freedom. We show that the method is robust to reducing the training data in space (number of sources) and time, which makes it potentially suitable for clinical applications.
\end{abstract}

\vspace{2pc}

%\submitto{\JNE}
\maketitle

\section{Introduction}

Prosthetic hardware has been significantly advanced in the last decade. However, the commercially available control interfaces embody paradigms developed decades ago. Bionic hands aim to provide much needed functionality to a broad range of users though they are limited by unintuitive and cumbersome control paradigms. Current commercial state-of-the-art systems rely on a two-electrode approach where each recording site senses the electromyographic signals (EMGs) of the underlying muscles. When either of the electrodes detects the EMG surpassing a given threshold, a pre-mapped direction of a prosthetic degree of freedom (DoF) is activated. This allows manipulation of only single DoFs at a time and requires an introduction of state changing triggers (e.g., a co-contraction, external sensor or a switch) to allow the control of another individual prosthetic DoF.

To improve current clinical systems, researchers have attempted various approaches in order to obtain and map sufficiently descriptive neural information into the prosthetic systems. Volitional motor command signals can be extracted from nerves (ENG) or more traditionally from muscles. While ENG approaches are only possible with invasive recordings, EMG recordings can be performed both non-invasively (surface EMG or sEMG) and intramuscularly (iEMG). Though more convenient, classic EMG does not provide direct access to individual motor neuron behavior, but rather to its filtered representation. An EMG electrode indeed records a spatiotemporal integration of electrical activity of many muscle fibers, creating a global signature of muscular activity \citep{merletti2004}. Accessing sources of motor commands requires solving an inverse problem that needs multiple observations (electrodes) and specialized processing. The main issue is that the number of sensors in commercial prosthetic systems is less than the number of sources. Due to these difficulties, most of currently available prosthesis controllers only use either global features of EMG signals for direct control, supervised EMG pattern classifiers to identify a few predefined movements \citep{erik2011, chowdhury2013} or regression-based approaches for estimating kinematics of few selected DoFs \citep{hahne2014}. While drawbacks of direct control are obvious, pattern recognition strategies are limited by a concept of motion execution that uses sequences of on-off sub-tasks, which means they can only control one sub-task at a time and at a fixed speed. For example, linear discriminant analysis \citep{englehart2003}, hidden Markov models \citep{chan2005}, support vector machines \citep{oskoei2008, kaufmann2010, geng2011}, and Bayesian learning \citep{young2013classification} are used with effective feature extraction methods for this purpose with very low classification error. However, these methods are challenged when concurrent movements are introduced. On the other hand, regression approaches do not compartmentalize the solution space to predefined motion classes, but rather intrinsically allow concurrent motions to be estimated. However, these approaches underperform when dealing with a large number of DoFs. Ultimately, due to poor functionality of unintuitive control schemes, the type, quality and precision of commercial upper-limb prostheses are currently limited and not well accepted by their users \citep{atkins1996, gaine1997, davidson2002, biddiss2007, hahne2014}.

Recent advances in multi-channel recording of neuromuscular signals and special surgical procedures along with modern signal processing methods have opened new opportunities to improve the control and use of hand prostheses. Targeted muscle reinnervation (TMR) has been used to restore neural pathways \citep{hijjawi2006, kuiken2007}, such that EMG recordings from the reinnervated muscle sites provide more degrees of freedom with higher signal to noise ratio \citep{hijjawi2006, zhou2007decoding}. Furthermore, signal decomposition methods like multiple input multiple output (MIMO) deconvolution \citep{yu1999} and independent component analysis \citep{hyvarinen1999fast, hyvarinen2000independent} are now applicable to the newly available high-density and high-sampling rate electromyography \citep{holobar2014blind, muceli2015accurate, chen2016, negro2016, andalib2018framework}. Although these approaches are still being refined for higher performance and real-time implementation, they are promising for providing direct access to underlying neural data by decomposing EMG signals into a high-dimensional set of individual motor neuron activities that represents the exact timings of motor neuron firings.

To extract the relevant information from such a diverse and high-dimensional data-set, one solution is to optimize this neural response with a possibly low dimensional representation where it is easier to investigate how the information related to different movements is represented. This approach is inspired by the working principles of the neuromuscular system that controls movements by activating muscles as integrators of the high dimensional motor neuron firings into low-dimensional continuous forces. From a signal processing perspective, the muscle system projects the high dimensional peripheral nervous system output into the subspace of movement trajectories. The decoding algorithm therefore needs to perform a dimensionality reduction projection, which may be achieved by metric learning \citep{xing2003distance}. A metric projection guarantees that (dis)similar neural responses in the high dimensional space (HDS) are projected close to (away from) each other in the low dimensional space (LDS), and that LDS remains continuous. By optimizing the proper weighting of individual dimensions, this approach learns to project the non-informative data dimensions into the LDS null space. Since the weighting values are not limited to zero-or-one contributions of dimensions, metric learning approximates the desired trajectory independently for each degree-of-freedom.

In this paper, we introduce a decoding approach for prosthetic hand control based on metric learning and apply the method to motor unit spike trains (MUSTs) obtained by blind source separation of high-density EMG signals recorded from the new muscle sites in an amputee after TMR procedure. The proposed approach exploits the spectral information in the high dimensional space of the neuro-muscular system and projects the relevant information to a low-dimensional subspace of continuous variable control signals of actuators of the prosthetic hand. These control signals are an estimation of the low-dimensional, task-level representation of movement commands by the peripheral nervous system (PNS) and the central nervous system (CNS) at a higher level. Here, we use this estimation to reconstruct the desired kinematics of concurrent motions. We believe that an inverse controller established in this manner, once trained, will provide the commands for an artificial hand to mimic the desired hand trajectories. This controller commands will define implicitly the projection space for the metric learning as desired signals.

A linear unsupervised metric projection followed by a rotation step will be introduced. The method will be tested as an offline controller using a rate based decoding approach to quantify muscle contraction from motor neuron discharge timings collected with a High-Density (HD) array of sEMG electrodes. The motor neuron discharge timings are obtained by an EMG decomposition algorithm and the accuracy of the proposed interface is analyzed on a transhumeral amputee who has undergone a TMR procedure. Progressively, up to three concurrent DoFs are tested.

\section{Methodology}
\subsection{Decomposition algorithm for multi-channel EMG signal}\label{decomp}

We propose a modification of the EMG decomposition algorithm introduced in \citet{negro2016} to better represent the total motor neuron activity for higher projection performance. The algorithm includes four major steps, namely (1) extending the EMG channels in time with delayed inputs to insert temporal redundancies, (2) whitening the extended measurements, to make them uncorrelated, (3) applying the independent component analysis (ICA) to regenerate the sequence of electrical pulses during muscle contractions, known as innervation pulse trains (IPTs), and (4) extracting MUSTs from IPTs by classification algorithms. The improvement mainly concerns the classification task in step (4) and as shown later with some experiments, it can significantly improve the decomposition task, and consequently the decoding performance.

The generation model of one channel of EMG signals can be described as a convolutive mixture of a series of delta functions, representing the discharge timings of the motor neurons \citep{holobar2007gradient, holobar2010experimental}. The impulse responses of the filters in this convolutive mixture are the motor unit action potentials (MUAPs) that have finite durations. This model is defined as
\begin{eqnarray}
\eqalign{
x_i\left(k\right)=\sum_{l=0}^{L-1}\sum_{j=1}^{n}{h_{ij}\left(l\right)s_j\left(k-l\right)}+n_i\left(k\right) \cr
\ i=1,\ldots,m,\qquad k=0,\ldots,D_r
},\label{eq:1}
\end{eqnarray}
where $m$ and $n$ are the number of EMG channels and active motor units, respectively, and $L$ is the finite duration of MUAPs, in samples. Here, $x_i(k)$ is the $k$-th sample of the $i$-th EMG channel for the duration of EMG recording $D_r$, in samples. $h_{ij}$ is the action potential of the $j$-th motor unit $(j=1,\ldots,n)$ as recorded at the $i$-th channel, $s_j(k)$ is the spike train of the $j$-th motor unit, and $n_i(k)$ is the additive noise at $i$-th channel.

Consider $\mathbf{x}\left(k\right)=\left[x_1\left(k\right),x_2\left(k\right),\ldots,x_m\left(k\right)\right]^\top,\mathbf{s}\left(k\right)=[s_1(k),s_2(k),\ldots,s_n(k)]^\top$, and $\mathbf{n}\left(k\right)=[n_1(k),n_2(k),\ldots,n_m(k)]^\top$ are the column vectors of the $k$-th sample of the $m$ observations, the $n$ sources, and the additive noise, respectively. \Eref{eq:1} can be rewritten in matrix form as
\begin{eqnarray}
\mathbf{x}\left(k\right)=\sum_{l=0}^{L-1}\mathbf{H}\left(l\right)\mathbf{s}\left(k-l\right)+\mathbf{n}\left(k\right).\label{eq:2}
\end{eqnarray}

\Eref{eq:2} implies that the set of unknown individual sources $\mathbf{s}$ is `mixed' in space and time using the $m\times n$ matrix $\mathbf{H}$ to produce the set of mixed observations $\mathbf{x}$. The goal of the blind signal separation here is to invert \eref{eq:2} effectively to find, from the recorded EMG channels $\mathbf{x}$, the largest number of decomposed sources of motor neuron spike trains $\mathbf{s}$. The `unmixing' matrix obtained by the BSS approach should decorrelate the spatio-temporal dependencies in the observations, a process also referred to as signal whitening.

Efforts have been made to solve the multi-channel EMG decomposition problem with latent component analysis approaches by simplifying assumptions. In recent works \citep{holobar2014blind, farina2015, negro2016, farina2017} the mixing matrix $\mathbf{H}$ is assumed to be constant for the recording duration. Furthermore, it is assumed that the identified sources are sparse and not fully correlated and the non-identified sources are modelled into the additive noise term $\mathbf{n}\left(k\right)$.

\cite{holobar2014blind}, \cite{farina2015}, and \cite{negro2016} suggest converting the convolutive mixture of sources into a linear instantaneous mixture by extending the sources to include the $n$ original sources as well as their delayed versions, with delays from 1 to the filter length $L$ samples. The $m$ original observations are also extended to maintain a greater number of observations than sources. The model then changes to
\begin{eqnarray}
\overline{\mathbf{x}}\left(k\right)=\overline{\mathbf{H}}\overline{\mathbf{s}}\left(k\right)+\overline{\mathbf{n}}\left(k\right),\label{eq:3}
\end{eqnarray}
where $\overline{\mathbf{x}}\left(k\right),\overline{\mathbf{s}}\left(k\right),\overline{\mathbf{n}}\left(k\right)$ are the extended observations, extended sources, and extended noise vectors, respectively. Notice the extended matrix $\overline{\mathbf{H}}$ is assumed to be fixed over time.

The linear instantaneous model of \eref{eq:3} is then inverted to recover the matrix of the extended sources by using only spatial whitening, since the temporal dependencies are now represented by spatial redundancies using source extension procedure.

A whitening matrix $\overline{\mathbf{W}}$, is obtained by imposing that the whitened extended observation matrix
\begin{eqnarray}
\overline{\mathbf{y}}\left(k\right)=\overline{\mathbf{W}}\overline{\mathbf{x}}\left(k\right), \label{eq:4}
\end{eqnarray}
has a covariance matrix equal to the identity for time lag zero, which is a valid assumption as shown in \cite{negro2016}. Then the whitening matrix is obtained as
\begin{eqnarray}
\overline{\mathbf{W}}=\mathbf{U}\mathbf{D}^\gamma\mathbf{U}^\top, \label{eq:5}
\end{eqnarray}
with $\gamma=-\frac{1}{2}$, and where $\mathbf{U}$ is given by the diagonalization matrix of the covariance matrix of the extended observation.

The spatial whitening is followed by a fixed-point optimization procedure \citep{hyvarinen1999fast}, as used in \cite{thomas2006time}, with a Gram-Schmidt orthogonalization step to increase the number of unique sources \citep{hyvarinen2000independent}. This procedure works with a cost function that maximizes the sparseness of the estimated sources \citep{negro2016}. The method extracts sources (series of motor neuron discharge timings) associated to individual motor neurons, as proven by the unique representation of the associated multi-channel surface action potentials \citep{farina2008detecting}. The estimated sources are trains of delta functions centered at the instant of motor neuron activation, with an amplitude that may vary due to the estimation process. This amplitude in the IPTs is a way to extract individual motor neuron discharge timings in the estimated sources. To detect large amplitude spikes, a K-means classification algorithm with two classes (spikes and non-spikes) has been previously used \citep{holobar2007gradient, negro2016}, which gives rise to a fixed detection threshold. This algorithm showed to be quite robust to avoid merged sources, but it could miss smaller peaks when the action potentials change in shape during a contraction in dynamic condition. In our work, missing motor neuron spike trains have a big effect on the output of the projection algorithm, as it determines erroneous dropouts in estimated movements. In contrast, the false detections will be twitches because they only occur in a few MUST channels and are very short, so they have less effect on the projection.

We modified the peak detection procedure by using a simple locally adaptive thresholding for each source channel, which compares each candidate with the surrounding values. We also considered a refractory period of 10 ms for motor neurons to limit false-positive detections and merged sources.

Finally, we computed the silhouette value (SIL) to identify the reliable sources. The SIL was defined as the difference between within- and between-cluster sums of points-to-centroid distances, normalized by the maximum of the two values. In this study, a source was qualified if SIL was greater than 0.8.

It is relevant to note that the decomposition method extracts the original sources as well as their delayed replicas, as defined by the extended source model of \eref{eq:3}.
\subsection{Metric learning for decoding} 
Metric learning is a projection framework that extends Fisher discriminant analysis \citep{mclachlan2004} to choose the best metric for the neural responses and its projections to sensorimotor integration. Metric learning allows selection of arbitrary real scalar weightings for each of the high dimensional inputs to provide required mappings to the subspace. Moreover, the projection is not restricted to be linear \citep{xing2003distance}. In our case, the output subspace is the space of commands for the hand prostheses. This is inspired by a common paradigm in brain-machine interfacing where the availability of a high dimensional space of motor neuron firings can produce adequate hand trajectories in three-dimensional space \citep{lebedev2005, gunduz2009}. Moreover, the result of using metric learning is that every input is weighted appropriately to represent the degrees of freedom for prostheses control. This paper employs principal component analysis (PCA), which is the simplest of the metric projections.

Our goal is to design an enhanced PCA projection that optimally maps relevant information from the high D-dimensional space of motor neuron discharge timings into the d-dimensional manifold $(d\ll D)$ of the control space of the kinematics DoFs, such that the space of desired trajectories in the DoF space are fully spanned, and the learned axes are sparse. The problem is formulated as follows. Given the $(n\times D)$ data matrix $\mathbf{X}$ of $n$ observations on $D$ dimensions measured about their sample means, assuming that $\mathbf{X}$ is full rank, PCA performs a singular value decomposition (SVD) of $\mathbf{X}=\mathbf{U\Sigma}\mathbf{W}^\top$, where $\mathbf{U}$, $\mathbf{W}$ are $\left(n\times D\right)$, $(D\times D)$ unitary matrices, respectively, and $\mathbf{\Sigma}$ is a $(D\times D)$ diagonal matrix with diagonal values $\sigma_i$ known as the singular values of $\mathbf{X}$. The columns of $\mathbf{W}$ are the eigenvectors of $\mathbf{X}^\top\mathbf{X}$ and they serve as the principal directions or axes of the sample covariance matrix $\mathbf{\Omega}$, and $\mathbf{\Sigma}$ gives the square roots of the eigenvalues of $\mathbf{X}^\top\mathbf{X}$, and hence the standard deviation of the principal components for $\mathbf{\Omega}$.

PCA uses $\mathbf{W}$ to map the input data $\mathbf{X}$ into
\begin{eqnarray}
\mathbf{XW}=\mathbf{U\Sigma}.\label{eq:6}
\end{eqnarray}

Looking at the mapping of a single observation may also be helpful for our purpose. PCA projects the observation vector $\mathbf{x}_i,\ i=1,2,\ldots,n$ to $\mathbf{y}_i=\mathbf{W}^\top\mathbf{x}_i$, where $\mathbf{x}_i^\top$ are the rows of $\mathbf{X}$. The mapping $\mathbf{x}_i\rightarrow\mathbf{y}_i$ is metric preserving in the sense that the Euclidean distance $d^2\left(\cdot,\cdot\right)$ between two samples $\mathbf{y}_i, \mathbf{y}_j$ on the projected space is equal to the Euclidean distance between $\mathbf{x}_i,\mathbf{x}_j$ in the original space
\begin{eqnarray}
\eqalign{
d^2\left(\mathbf{y}_i,\mathbf{y}_j\right)&=\left(\mathbf{y}_i-\mathbf{y}_j\right)^\top\left(\mathbf{y}_i-\mathbf{y}_j\right)\cr
&=\left(\mathbf{W}^\top\mathbf{x}_i-\mathbf{W}^\top\mathbf{x}_j\right)^\top\left(\mathbf{W}^\top\mathbf{x}_i-\mathbf{W}^\top\mathbf{x}_j\right)\cr
&={\left(\mathbf{x}_i-\mathbf{x}_j\right)}^\top\mathbf{WW}^\top\left(\mathbf{x}_i-\mathbf{x}_j\right)\cr
&=d^2\left(\mathbf{x}_i,\mathbf{x}_j\right),
}\label{eq:7}
\end{eqnarray}
as the columns of $\mathbf{W}$ are orthonormal, hence $\mathbf{W}\mathbf{W}^\top=\mathbf{I}$.

It can be shown that the mapping to best (in the projected power sense) control $d$ degrees of freedom is formed by the first $d$ eigenvectors, i.e., columns of $\mathbf{W}$, noted as $\mathbf{W}_d$. In this case, the distances between points on the projected space are an approximation to the Euclidean distance. Furthermore, SVD with the first $d$ components is an estimation to the input data
\begin{eqnarray}
\mathbf{X}\approx\mathbf{U}_d\mathbf{\Sigma}_d\mathbf{W}_d^\top=\mathbf{S}_d\mathbf{W}_d^\top, \label{eq:8}
\end{eqnarray}
where the raw scores are noted by $\mathbf{S}_d$.
 
Since PCA is an unsupervised technique, it is necessary to find a way of interpreting the projection output as to associate each principal component to each prosthetic DoF. The PCA subspace and the prosthetic's motor control space are both orthogonal and selected of the same dimension. However, the two coordinate systems are not necessarily the same because the PCA eigenvectors are solely defined by the data. Hence, a $(d\times d)$ orthogonal rotation $\mathbf{R}$ is needed to best align the principal directions with the external basis of DoFs. Moreover, the external basis coordinates implemented by the motors are the canonical sparse basis of $\mathds{R}^n$, while the PCA eigenvectors are not sparse and are ordered by projected variance. Therefore, we sought to find the rotation in the $d$-dimensional principal component subspace that makes the projection matrix as sparse as possible, instead of preserving the maximum power in the early principal directions as PCA does.
\subsection{Rotating the projection subspace}
The orthogonal rotation $\mathbf{R}\ (d\times d)$, given by any `simplicity' criterion \citep{harman1976}, like VARIMAX \citep{kaiser1958}, may be used to rotate the axes within the $d$-dimensional subspace. The simplicity function achieves some sparsification of the projection matrix. As it is shown later, the sparse eigenvectors redistribute the power among the rotated principal components, which determines a more even distribution compared with before-rotation power distribution in PCA subspace.

To formulate the rotation of principal directions (eigenvectors), and the raw scores, we plug the identity matrix $\mathbf{R}\mathbf{R}^\top$ into \eref{eq:8}:
\begin{eqnarray}
\mathbf{X}\approx\mathbf{U}_d\mathbf{\Sigma}_d{\left(\mathbf{R}\mathbf{R}^\top\right)}\mathbf{W}_d^\top=\left(\mathbf{U}_d\mathbf{\Sigma}_d\mathbf{R}\right)\left(\mathbf{W}_d\mathbf{R}\right)^\top=\mathbf{S}_{rot}{\mathbf{W}_{rot}}^\top. \label{eq:9}
\end{eqnarray}

From \eref{eq:9}, it is clear that rotated eigenvectors $\mathbf{W}_{rot}\ (D\times d)$ are still orthogonal as
\begin{eqnarray}
{\mathbf{W}_{rot}}^\top\mathbf{W}_{rot}=\ \mathbf{R}^\top{\mathbf{W}_d}^\top\mathbf{W}_d\mathbf{R}=\mathbf{I}_d.\label{eq:10}
\end{eqnarray}

\begin{figure}
\centering
\includegraphics[scale=0.7]{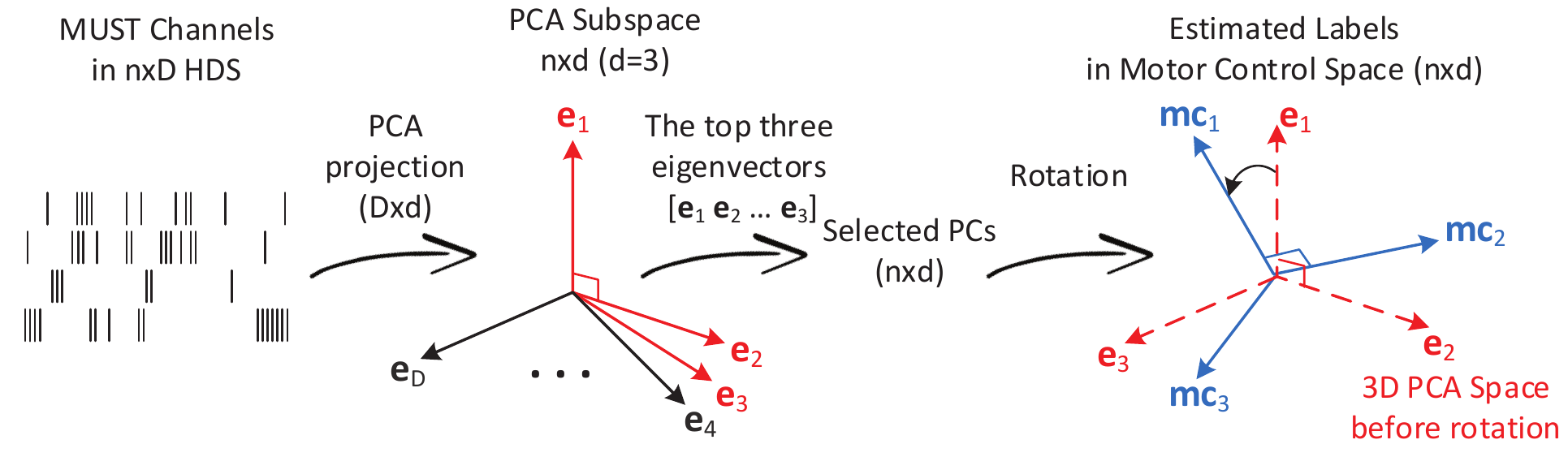}
\caption{PCA projection and rotation of PCA subspace (red axes) to align with the orthogonal motor control space (blue axes). For better visualization, a three-dimensional subspace is considered here.}
         \label{fig:1}
\end{figure}

We are interested in this framework because the rotated axes of the subspace are orthogonal, which is required when we aim to align the orthogonal projected subspace with the orthogonal space of DoFs. This idea is shown in figure \ref{fig:1}.

Here, we applied the VARIMAX orthogonal rotation $\mathbf{R}_{VARIMAX}$ \citep{kaiser1958}
\begin{eqnarray}
\mathbf{R}_{VARIMAX}=\argmaxAA_\mathbf{R} {\left[\frac{1}{d}\sum_{j=1}^{n}\sum_{i=1}^{d}{(\mathbf{W}_d\mathbf{R})}_{ij}^4-\sum_{j=1}^{n}\left(\frac{1}{d}\sum_{i=1}^{d}{(\mathbf{W}_d\mathbf{R})}_{ij}^2\right)^2\right]},\label{eq:11}
\end{eqnarray}
where $\mathbf{W}_d\mathbf{R}$ represents the rotated PCA loadings $\mathbf{W}_{rot}$. Because the term in the square brackets are proportional to the variance of squared PCA loadings for each rotated direction and the original loadings are necessarily between $-1$ and 1, the VARIMAX criterion tends to drive squared loadings toward the extremes of the range $0$ to 1. This means the loadings are pushed toward $-1$, 0 or 1, and away from intermediate values, as required by the simplicity criteria \citep{harman1976}.
\subsection{Experimental setup}
\subsubsection{Participant} A 51 years old male with 10 years of left transhumeral amputation volunteered to participate in the study. He underwent a TMR procedure 52 months before the experiments. The TMR involved surgical redirection of the medianus, ulnaris, and radialis nerves into three new muscles above the amputation, namely the brachialis, caput breve bicipitis, and caput laterale tricipitis muscles, respectively. The experimental protocol and the informed consent form for the experiment were approved by the ethics committee \textquotedblleft Ethikkommission der Medizinischen Universit\"{a}t Wien\textquotedblright{} (approval number 1234/2015).
\subsubsection{Participant task and testing} The movements involve one, two, or three DoF(s) of elbow flexion, wrist pronation, wrist flexion, and wrist extension. To perform these tasks, the subject followed ramp cues of 3-s duration for each side, peaking at 100\% of the maximum voluntary contraction (MVC). Resting periods were allowed between trials as chosen by the subject.

During the experiment, the subject was instructed to perform a series of mirror movements, with the able limb kinematics recorded as a reference using a motion capture system (OQUS 300+, Qualisys AB, Gothenburg, Sweden) and a set of retro-reflective markers at 256-Hz sampling frequency. High-density EMG was recorded using three surface grids of 64 electrodes and sampled at 2048 Hz with 12-bit precision (EMG-USB2, OT Bioelettronica, Torino, Italy). The EMG signals were then extended with temporal delayed version of the original channels (5~samples), whitened and decomposed using the blind source separation methodology, as described above. The extended extracted sources represent temporal redundancy exploited for temporal deconvolution of EMG electrodes and sum up the total number of delayed MUSTs to 251, 71, 244, and 247 in one, two, and three (two cases) DoF(s), respectively, as reported in table~\ref{tab:1}.
\subsubsection{Projection training and performance evaluation} The input to the PCA algorithm was built from spike counts (50-ms bin size) on every delayed MUST channel in the first two trials, and the projection was performed to a 12-dimensional data manifold, which provided enough degrees of freedom for subspace of movement trajectories. The VARIMAX rotation was then applied and the resulting projection matrix was used to estimate joint angles. We used the first two trials to find the active principal components for each DoF as well as the desired direction. Furthermore, since PCA does not provide the correct amplitude of the projected movement because of an un-modeled gain in the translation of motor activity to movement, we extracted the gain during the first trial to calibrate the rest of the trials. The third trial was only used for test purposes.

From the experiments, we observe that the projected angles were not as smooth as the recorded angles, which was expected because PCA is an instantaneous statistical projection. Hand movements are intrinsically low-pass filtered by the muscle dynamics, which was not modeled in our projection framework. To compensate for this, we pre-processed (low-pass filter) the input binned MUST channels to remove the ripples at the output without bringing large latency in the projection. The filter used was a first-order IIR with a cut-off frequency of 1 Hz, obtained from the analysis of the MUST data spectrum. The filter had a $-$45dB stop-band attenuation at 3 Hz.

The performance on the train\slash test trials was quantified in terms of multivariate $R^2$ index \citep{d2006}, that provides the ratio of total variation of movements captured by the estimation, with a maximal value of one for perfect estimation.

\section{Results}
\subsection{Projection performance: estimated vs. recorded kinematics}
\begin{figure}
\centering
\includegraphics[scale=0.65]{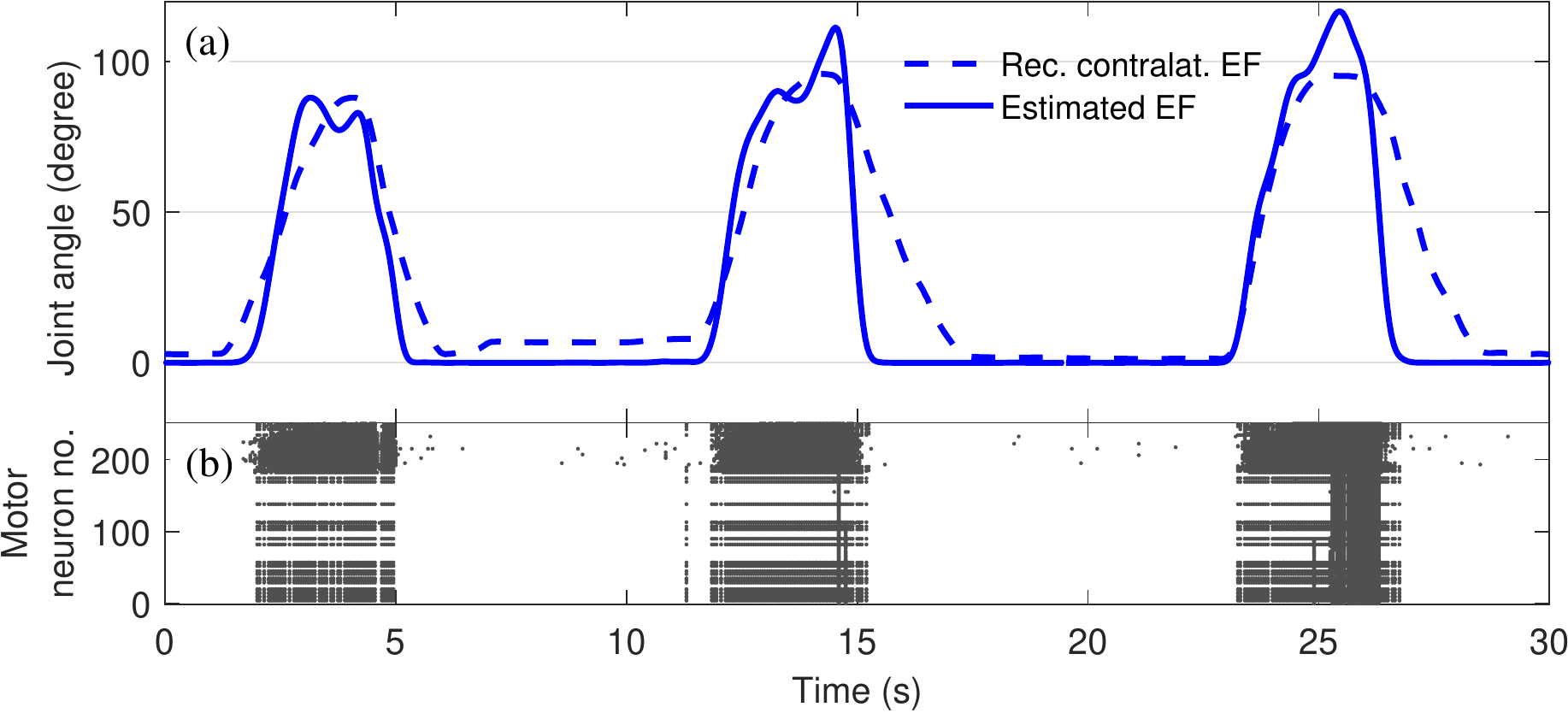}
\caption{(a) Decoding performance over a simple (one degree of freedom) movement, that is elbow flexion (EF). The estimated joint's angle trajectory (solid line) is compared with recorded contralateral (Rec. contralat.) elbow flexion on the other side (dotted line). (b) The raster plot of extended motor neuorn spike trains is also depicted for comparing.}
         \label{fig:2}
\end{figure}

\begin{figure}[!ht]
\centering
\includegraphics[scale=0.65]{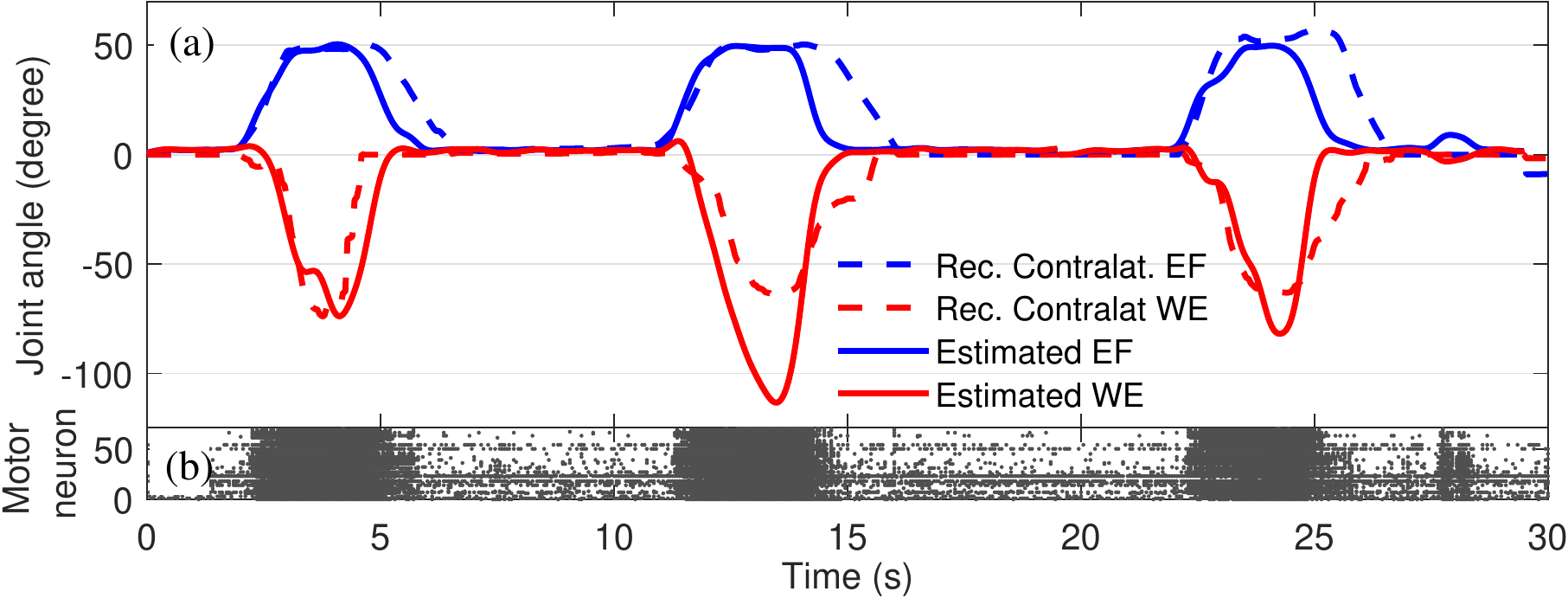}
\caption{(a) Decoding performance over two degrees composite movement with concurrent elbow flexion (EF) and wrist extension (WE) with (b) the raster plot of extended motor neuron spike trains in the bottom for comparing.}
         \label{fig:3}
\end{figure}

\begin{figure}[!ht]
\centering
\includegraphics[scale=0.65]{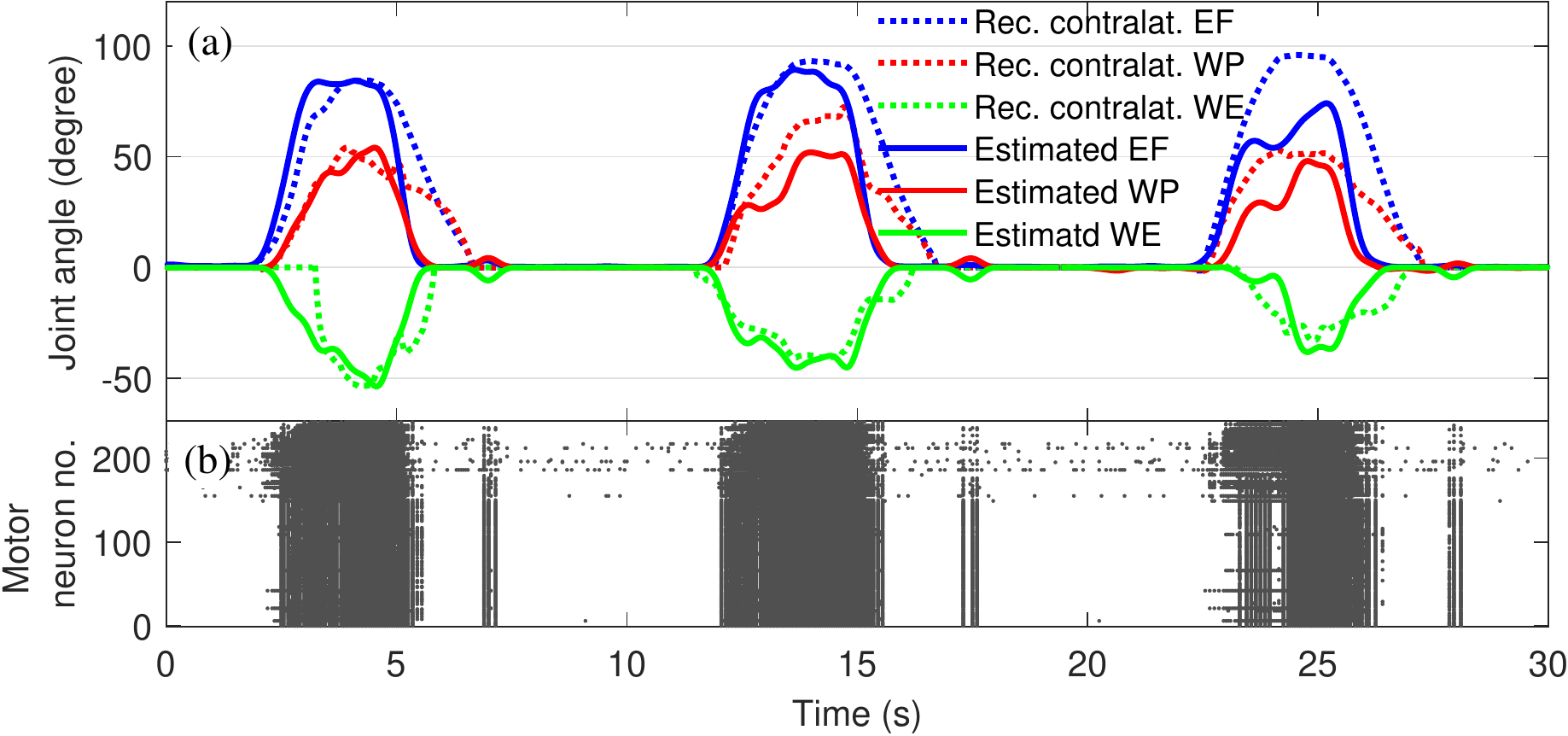}
\caption{(a) Decoding performance over three degrees composite movement with concurrent elbow flexion (EF), wrist pronation (WP), and wrist extension (WE) with (b) the raster plot of extended motor neuron spike trains in the bottom for comparing.}
         \label{fig:4}
\end{figure}

\begin{figure}[!ht]
\centering
\includegraphics[scale=0.65]{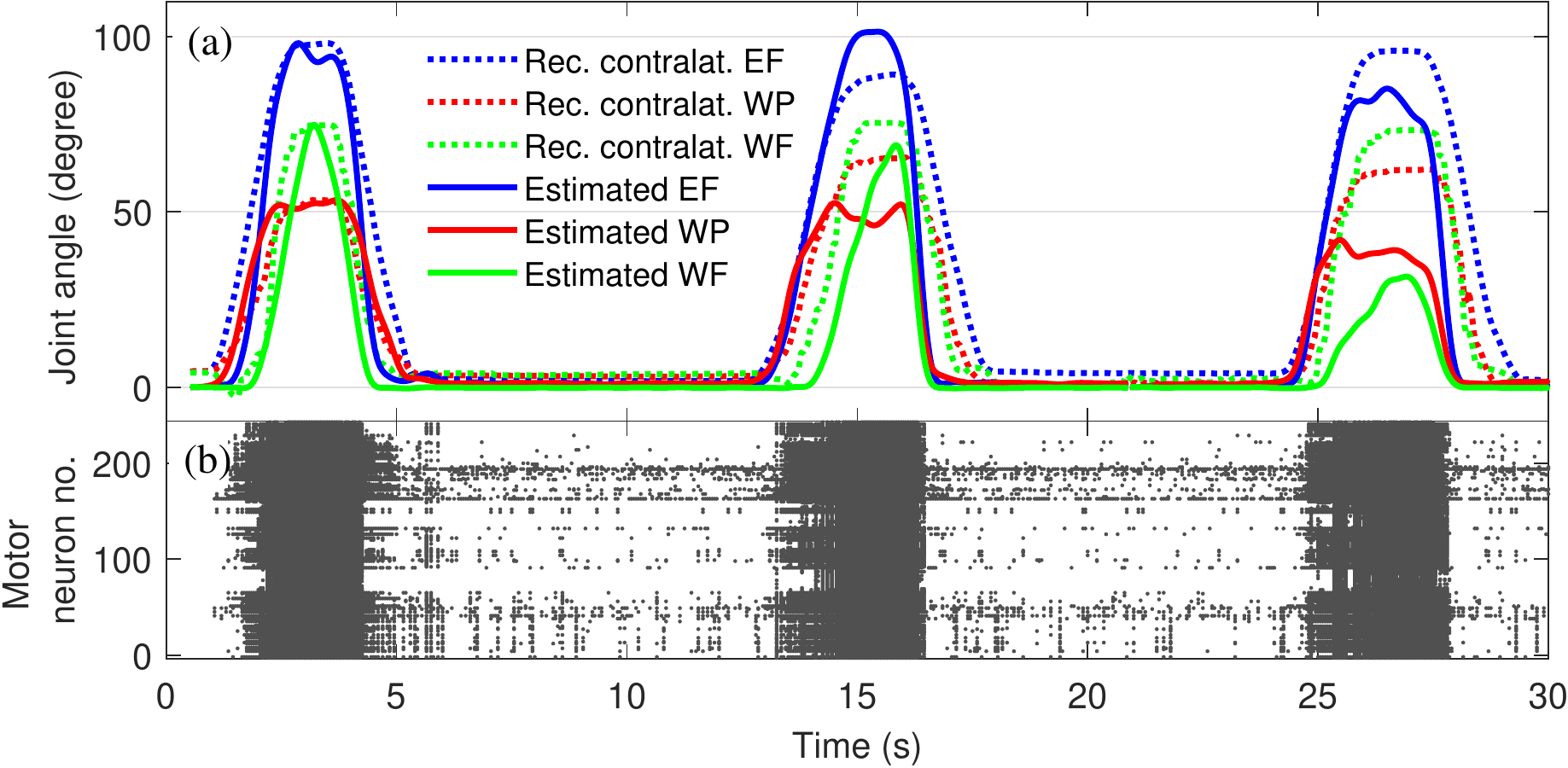}
\caption{(a) Decoding performance over three DoFs composite movements of elbow flexion (EF), wrist pronation (WP) and wrist flexion (WF) with (b) the raster plot of extended motor neuron spike trains in the bottom for comparing.}
         \label{fig:5}
\end{figure}

Figures \ref{fig:2}-\ref{fig:5} show the analysis performed on three similar trials with one, two and three (two cases) degree(s) of freedom, respectively. The raster plot of the corresponding motor neuron discharge timings including the delayed replicas extracted by decomposition of EMG signals are also shown in each case. The firing patterns are synchronized with the movement cues for comparisons.
 
Table~\ref{tab:1} provides the type of movement(s) and the number of decomposed MUST channels (extended motor neuron spike trains) in each experiment, along with the estimated $R^2$ values. These values represent an underestimation of the actual performance, because there is no guarantee that the mirror movements from the contralateral side were exactly the same as the phantom limb movements. Nonetheless, from the quantitative assessment presented in table~\ref{tab:1}, we can conclude that the projection generalized well to the test set. Furthermore, the variability in the training set was as high as that of the test set that is due to the small amount of data and the unsupervised nature of the algorithm (no overfitting).

\begin{table}
\caption{\label{tab:1}Estimation performance for train and test sets in terms of $R^2$ index.}
\footnotesize
\begin{tabular}{@{}lclcc}
\br
Case No.\,: No. of DoF(s)& No. of MUST Ch.$^{a}$ & Type of Movement & Train Set & Test Set\\
\mr
Case 1: 1 DoF (Figure \ref{fig:2}) & 251 & EF\,$^{b}$ & 0.81 & 0.77\\
Case 2: 2 DoFs (Figure \ref{fig:3}) & 71 & EF, WE\,$^{c}$ & 0.62 & 0.72\\
Case 3: 3 DoFs (Figure \ref{fig:4}) & 244 & EF, WP\,$^{d}$, WE & 0.82 & 0.65\\
Case 4: 3 DoFs (Figure \ref{fig:5}) & 247 & EF, WP, WF\,$^{e}$ & 0.87 & 0.73\\
\br
\end{tabular}\\
$^{a}$\ These are the `extended' motor neuron spike trains, as described in subsection \ref{decomp}\\
$^{b}$\,Elbow Flexion; $^{c}$\,Wrist Extension; $^{d}$\,Wrist Pronation; $^{e}$\,Wrist Flexion.
\end{table}
\normalsize

Further detailed analysis can be drawn from the figures after comparing the mirror movements with each other and with the motor neurons activities. Notice when the actual labels are not available due to amputation, considering the discharge timings can also reveal some information regarding the quality of estimation. Figure \ref{fig:5} illustrates the projection performance on a three-dimensional composite movement with concurrently active degrees of freedom including elbow flexion (EF) in blue, wrist pronation (WP) in red, and wrist flexion (WF) in green. The dotted lines represent the recorded (Rec.) joint angles of kinematics of the contralateral (contralat.) side of the subject, while the solid lines depict the first three principal components after rotation and proper calibration, as explained earlier. The figure also represents the raster plot of the corresponding extended motor neuron discharge timings. 

As shown in the figures, rising time detection in the third trial (test set) is generally as good as that of training trials. However, a major source of discrepancy is the inaccuracy in amplitude estimation during movements or other abrupt amplitude drops when returning to the original position. In this regard, we analyzed the source of error and we noticed this was partially due to the noise in the MUST channels (extracted sources). 

\subsection{The effect of rotation: eigen-structure before/after rotation}
\begin{figure}
\centering
\includegraphics[scale=0.45]{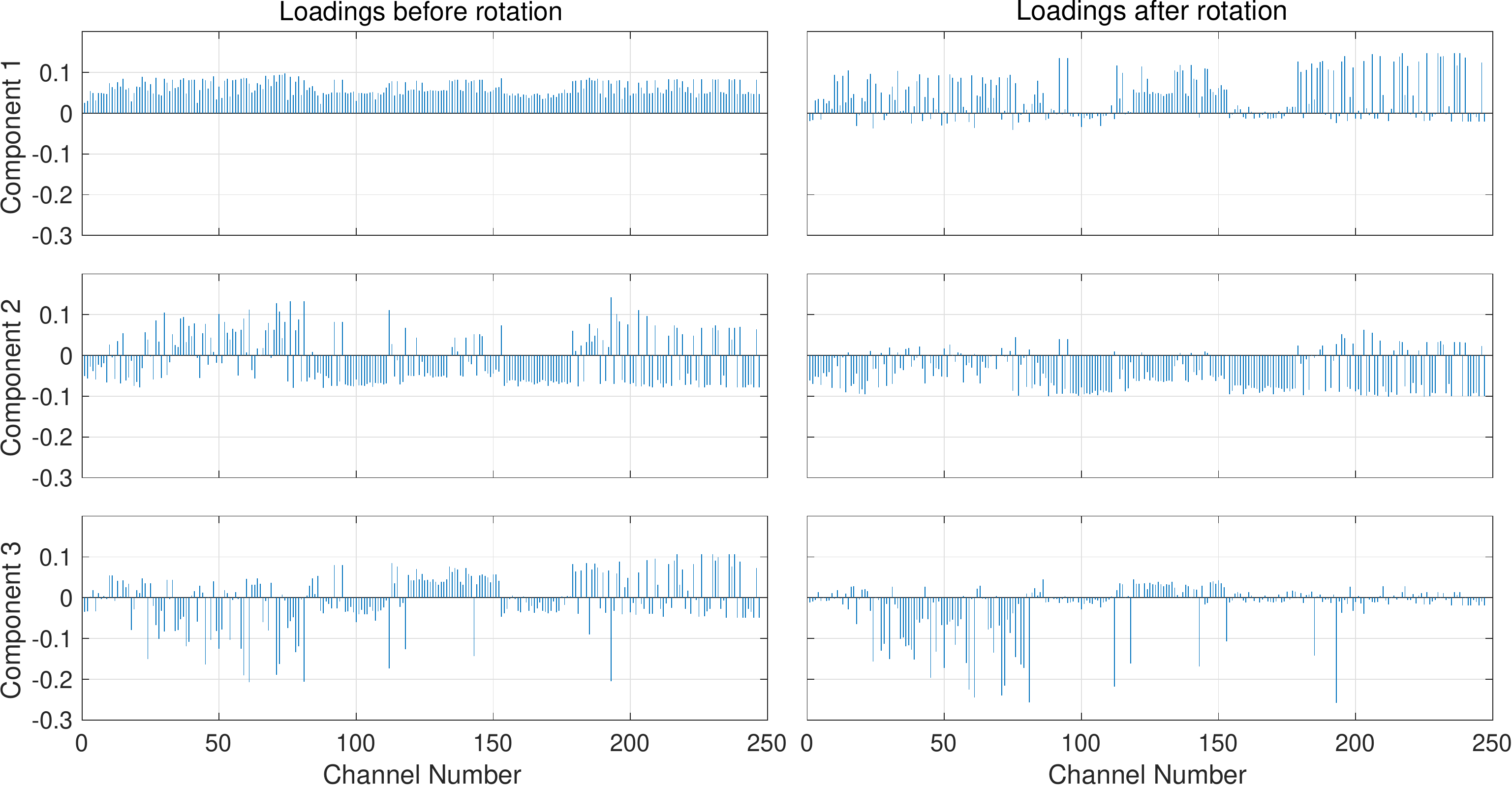}
\caption{Three first loadings vectors before (left panel) and after (right panel) VARIMAX rotation.}
         \label{fig:6}
\end{figure}

Studying the eigen-structure of the projection matrix before and after rotation helps to understand how VARIMAX rotation applies to our problem. This method maximizes the sum of the variances of squared loadings (i.e., the squared correlations between variables and factors). Figure \ref{fig:6} illustrates some of the primary loadings before rotation (left panel) compared with loadings after rotation (right panel). It can be seen that some of the weights are moved toward zero, and some of them are maximized, in a way that the whole projection matrix becomes as sparse as possible. Thus, VARIMAX transforms the projection matrix in the well-known simple structure \citep{harman1976} by detecting the MUST inputs that mostly contribute to control each DoF.

\begin{figure}
\centering
\includegraphics[scale=0.4]{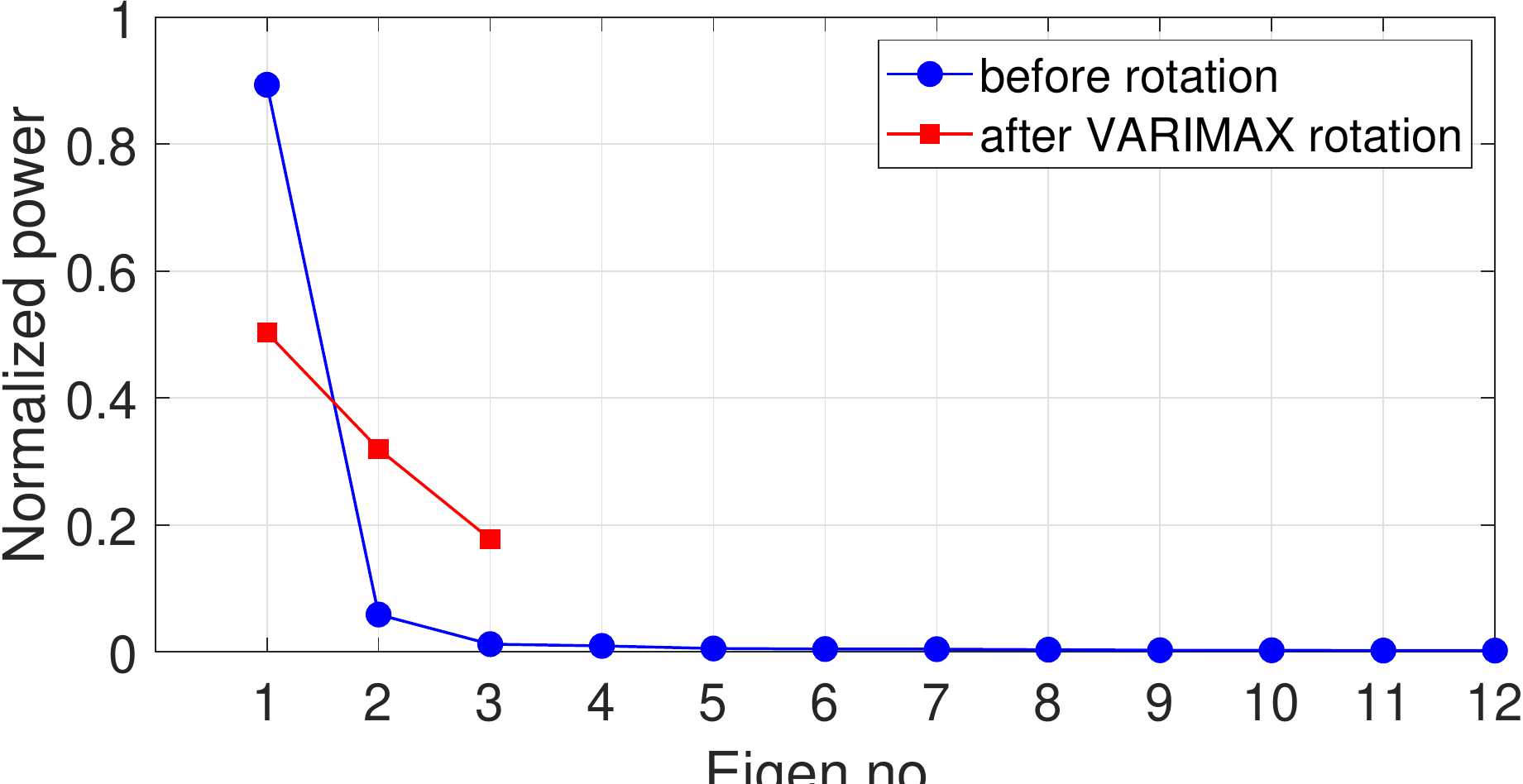}
\caption{PCA eigenspectrum in blue versus power in the first three dimensions after VARIMAX rotation.}
         \label{fig:7}
\end{figure}

Inspecting the rotation matrix\\
$\mathbf{R}_{VARIMAX} =
\left[
\begin{array}{ccc}
0.72855&-0.54932&-0.40922\\0.32689&0.80381&-0.49703\\0.60196&0.22834&0.76518\\
\end{array}
\right]$,\\
we observe that the rotation compromises the sparseness in the first eigen direction, but distributes better the remaining energy in the second and third directions. This is also seen in figure \ref{fig:7}. The eigenvalues of the projection (blue) are compared with the power in the first three dimensions after rotation. While the total variance stored in the top three dimensions is unchanged after rotation, it is redistributed amongst the rotated dimensions more evenly than before rotation.

\subsection{Decoding performance under implementation constraints}
We showed that simultaneous and proportional estimation of concurrent movements with up to three DoFs is possible, if we can read continuously from a large number of motor neurons over space and time. This requires a large number of EMG electrodes and a complex EMG decomposition algorithm applied to the whole set of electrodes. However, because of hardware constraints in clinical applications, such as cost, bandwidth and power consumption, as well as time constraint in online control, we may have to either reduce the number of EMG electrodes, which consequently reduces the number of extracted motor unit spike trains, or decompose subsets of EMG electrodes at a time, with a switching between them such that it covers all the motor neurons' activities. The former means we can only read from a subset of motor neurons after decomposition (space limitation), while the latter imposes switching between motor neurons that prevents access to every instance of motor neuron discharge timings (time limitation). Indeed, the space-time sub sampling of EMG electrodes is directly extended to space-time sub sampling of motor neurons' activities over MUST channels. So, it is of practical importance to further investigate how the performance of the decoding algorithm is affected by these implementation constraints. Here, we analyze and quantify the effect of reducing the number of MUST channels and then propose an approach that we call time-multiplexing to tackle the issue by trading space versus time resolution. To do so, we simulate and quantify the effect of switching when reading from binned spike trains.
\subsubsection{Reduced data-set analysis}
We investigated the performance of the algorithm with reduced subsets of 8, 16, 32, 48, 96, and 192 MUST channels, randomly selected from the set of extended motor neuron spike trains. For case 2, only 71 extended sources were extracted. The analysis was performed across 50 independent runs. The mean and variance of performance on reduced data-sets for case 4 (Ref. table \ref{tab:1}, three DoFs, EF, WP, WF) are reported in table \ref{tab:2} and is compared with that of the full set of motor neurons.

\begin{table}
\caption{\label{tab:2}The mean and variance of $R^2$ values over train and test sets for 3-DoF movements (case 4).}
\footnotesize
\begin{tabular}{@{}ccc}
\br
No. of MUST Ch. & Train Set & Test Set\\
\mr
8 & 0.8069$\pm$0.0060 & 0.6523$\pm$0.0077\\
16 & 0.8264$\pm$0.0037 & 0.6846$\pm$0.0033\\
32 & 0.8531$\pm$0.0010 & 0.7070$\pm$0.0013\\
48 & 0.8552$\pm$0.0011 & 0.7174$\pm$0.0010\\
96 & 0.8671$\pm$0.00003 & 0.7236$\pm$0.0003\\
192 & 0.8665$\pm$0.00001 & 0.7276$\pm$0.0006\\
247 & 0.8675 & 0.7292\\
\br
\end{tabular}\\
\end{table}
\normalsize

\begin{figure*}[!t]
    \centering
    \begin{subfigure}[]
        \centering
        \includegraphics[scale=0.4]{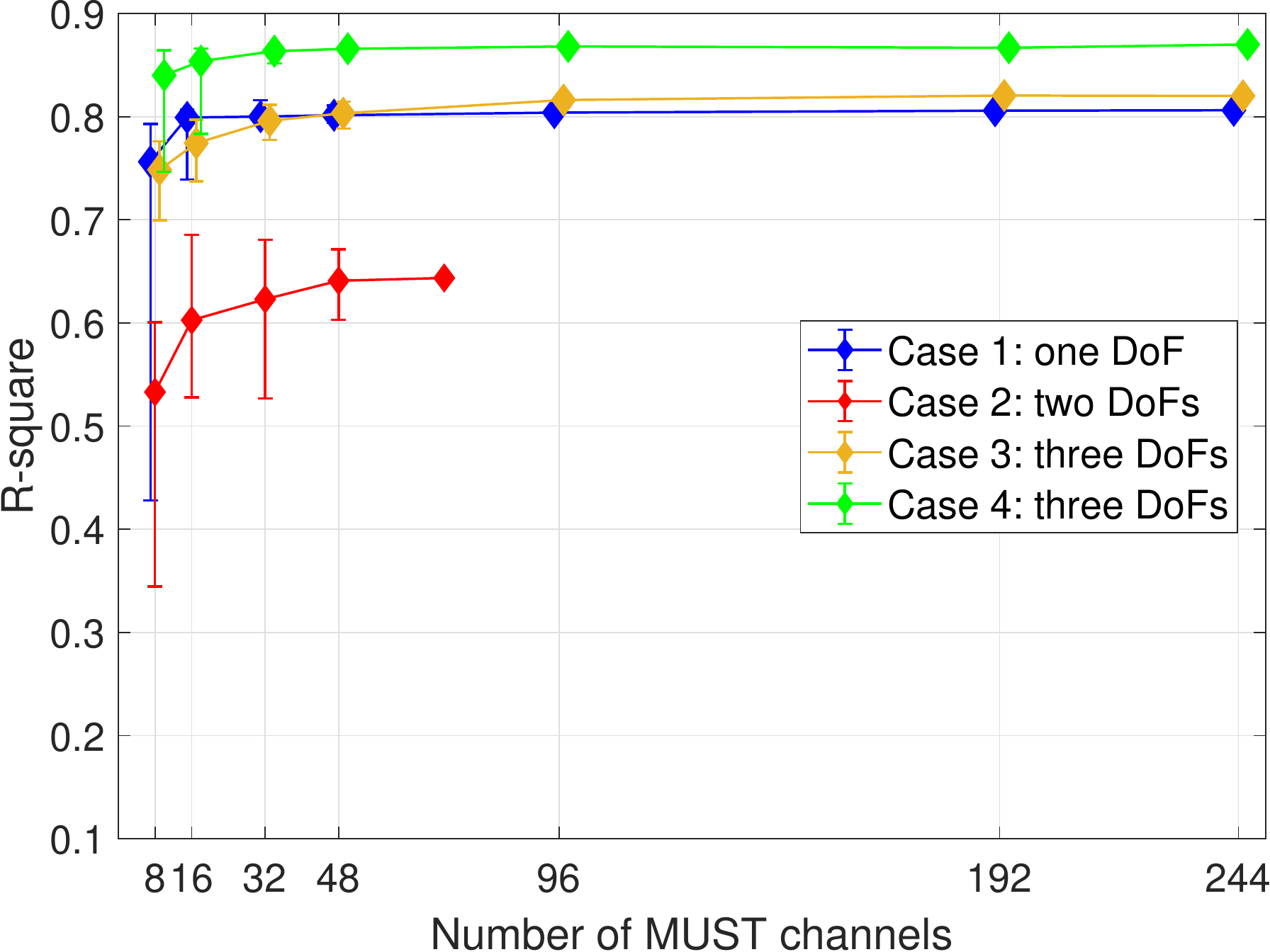}
        %\caption{}
    \end{subfigure}%
    ~ 
    \begin{subfigure}[]
        \centering
        \includegraphics[scale=0.4]{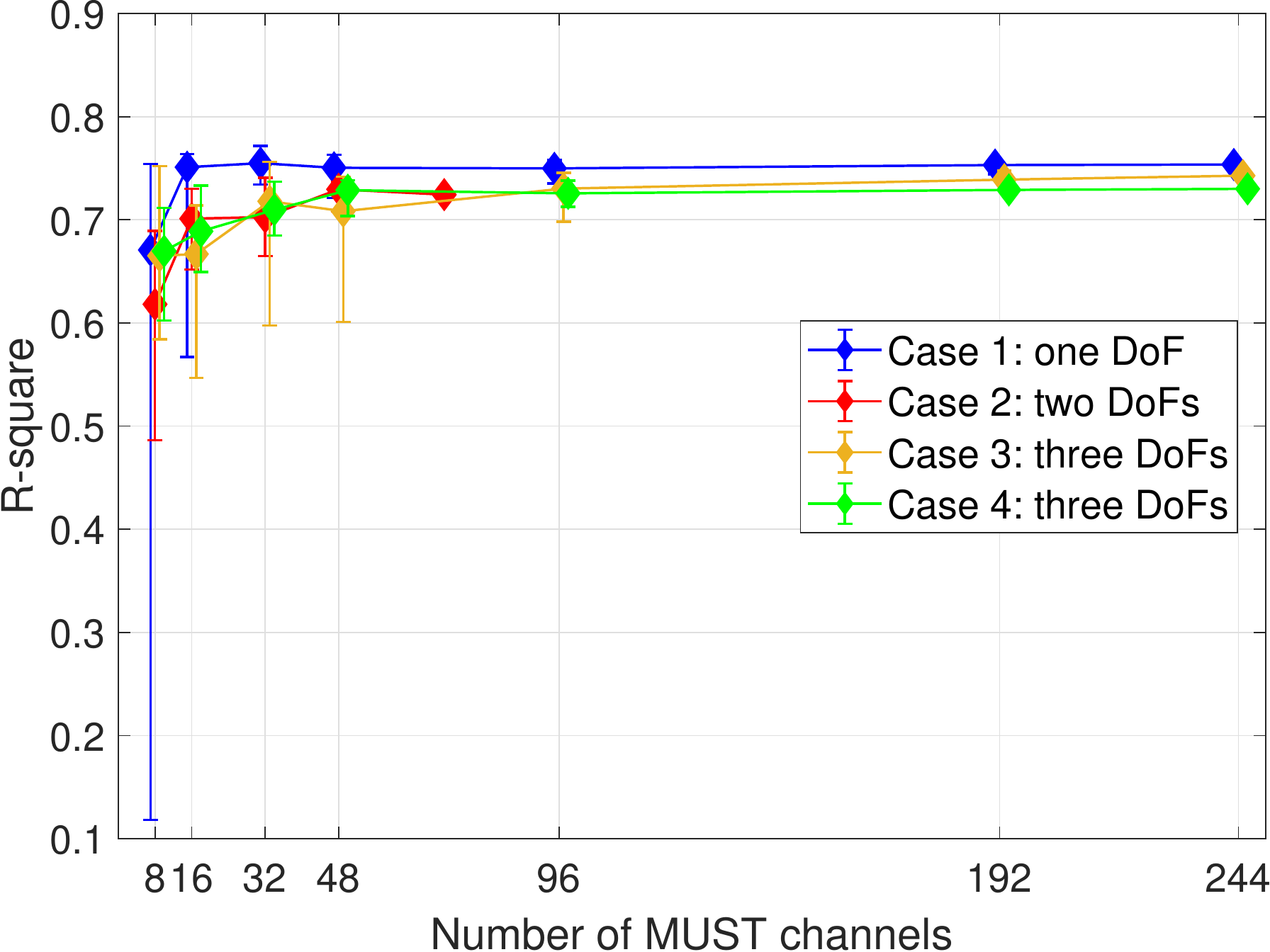}
        %\caption{}
    \end{subfigure}
    \caption{Reduced data-sets analysis. 50 independent runs, with randomly selected MUST channels, are performed over (a) train-set and (b) test-set. The median and 25\slash75 percentiles of $R^2$ index are depicted.}
\label{fig:8}
\end{figure*}

Similar analysis was performed for all other cases, with median and 25\slash75 percentiles reported in figure \ref{fig:8}. As expected, the performance slightly decreased with reducing the number of MUSTs used for the projection. Notice 32 is at the knee of performance where the variance is small enough for decoding reliability. This also suggests the minimum size of random subsets when switching between MUST channels in the time-multiplexing algorithm introduced in the next subsubsection. The performance differences on test-set are quite small when using more than approximately 96 MUST channels, when the performance saturates. Since in the decomposition algorithm we used an extension factor of 5 samples, 96 MUST channels represent the neural activity of about 20 motor neurons after removing some poor estimations. According to our observations, performance outliers ($R^2$ smaller than lower adjacent value) appear when the number of MUSTs is less than 96 and considerably increases when using less than about 32 MUST channels. 

\subsubsection{Time-multiplexing analysis}

To simulate time-multiplexing, the input data matrix was reconstructed by updating only one block of the input at a time (with few samples), in a periodic or random manner, while the other blocks keep their past values. The goal is to preserve as much as possible of the projection fidelity in one channel over time. Here are the steps for reconstruction

\begin{enumerate}
  \item We read from a random subset of minimum size of 32 MUST channels (known from reduced data-set analysis), without replacement, for a few samples of 50ms time bins. The time length of this block of data (related to the number of samples) is given by $T_B$.
  \item We then switch to a different random subset, as many times as required to visit all or a predetermined number of channels that we know, from reduced data-set analysis, is large enough to capture all the motor activity. The number of switchings is given by $N$.
  \item Once we fully scan all the MUST channels, we revisit the initial set to update the first block in a revisiting time $T_R$ that equals $N$ by $T_B$.
\end{enumerate}

We analyzed the effect of time-multiplexing in case 4 with different switching parameters (setups). Table \ref{tab:3} summarizes the train-set and test-set performances of switching across 50 independent runs, and compare those with no switching scenario on reduced or full data-sets.

\begin{table}
\caption{\label{tab:3}The $R^2$ values for no switching scenario versus switching with different parameters.}
\footnotesize
\begin{tabular}{@{}ccc}
\br
Setup No. & Train Set & Test Set\\
\mr
1 & 0.8442$\pm$0.00019 & 0.6758$\pm$0.00019\\
\textbf{2} & \textbf{0.8811$\pm$0.00013} & \textbf{0.7364$\pm$0.00063}\\
3 & 0.8630$\pm$0.00026 & 0.7124$\pm$0.00039\\
4 & 0.8531$\pm$0.0010 & 0.7070$\pm$0.0013\\
5 & 0.8671$\pm$0.00003 & 0.7236$\pm$0.0003\\
6 & 0.8675 & 0.7292\\
\br
\end{tabular}\\
Setup 1: reading 32 out of 224 MUST channels, $T_B=200$ms (4 samples),\ $N=7$,\ $T_R=1400$ms\\
Setup 2: reading 32 out of 96 MUST channels, $T_B=100$ms (2 samples), $N=3$, $T_R=300$ms\\
Setup 3: reading 32 out of 96 MUST channels, $T_B=50$ms (1 sample), $N=3$, $T_R=150$ms\\
Setup 4: reduced set of 32 MUST channels without switching\\
Setup 5: reduced set of 96 MUST channels without switching\\
Setup 6: Using all the information (all MUST channels at all times)\\
\end{table}
\normalsize

\begin{figure}
\centering
\includegraphics[scale=0.35]{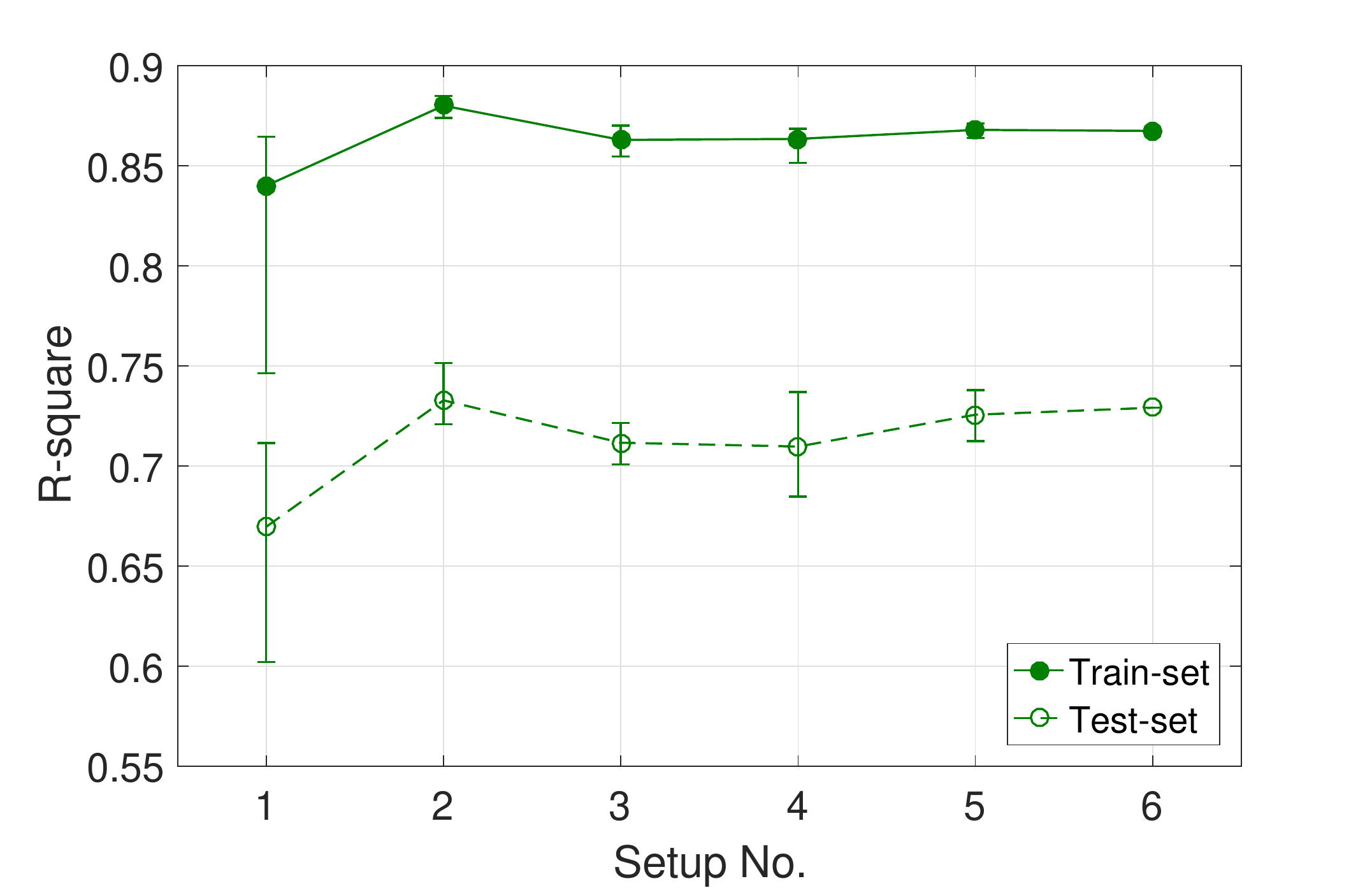}
\caption{Time-multiplexing analysis. For case 4 (three DoFs), different setups are tested across 50 independent runs over train-set (solid line) and test set (dotted line). The median and 25\slash75 percentiles of $R^2$ index are depicted.}
         \label{fig:9}
\end{figure}

Figure \ref{fig:9} compares the median and 25\slash75 percentiles. From this analysis, we can conclude that reading one block of 32 MUST channels with two samples at a time (setup 2) is the best trade-off between time and space resolutions. This setup even outperforms the setup using the whole number of MUST channels at all times, not only on average but in 25\slash75 percentiles, which is very interesting. A possible reason that needs more investigation is based on the natural time-variant activation model of muscle fibers when trying to avoid fatigue by changing the rate of excitation of muscle fibers. Our projection method can partially handle this time-variant behavior, because the input includes many channels of motor neuron spike trains. So, all we need is to find a gain factor for the amplitudes. However, time-multiplexing (with appropriate parameters) makes our projection more robust against this un-modeled characteristic of natural motor control, because it keeps an input block fixed over time before revisiting that block again. Notice that there are no considerable delays or distortion in time after doing time-multiplexing.

\subsection{Comparing adaptive vs. fixed thresholding} 
\begin{figure}
\centering
\includegraphics[scale=0.65]{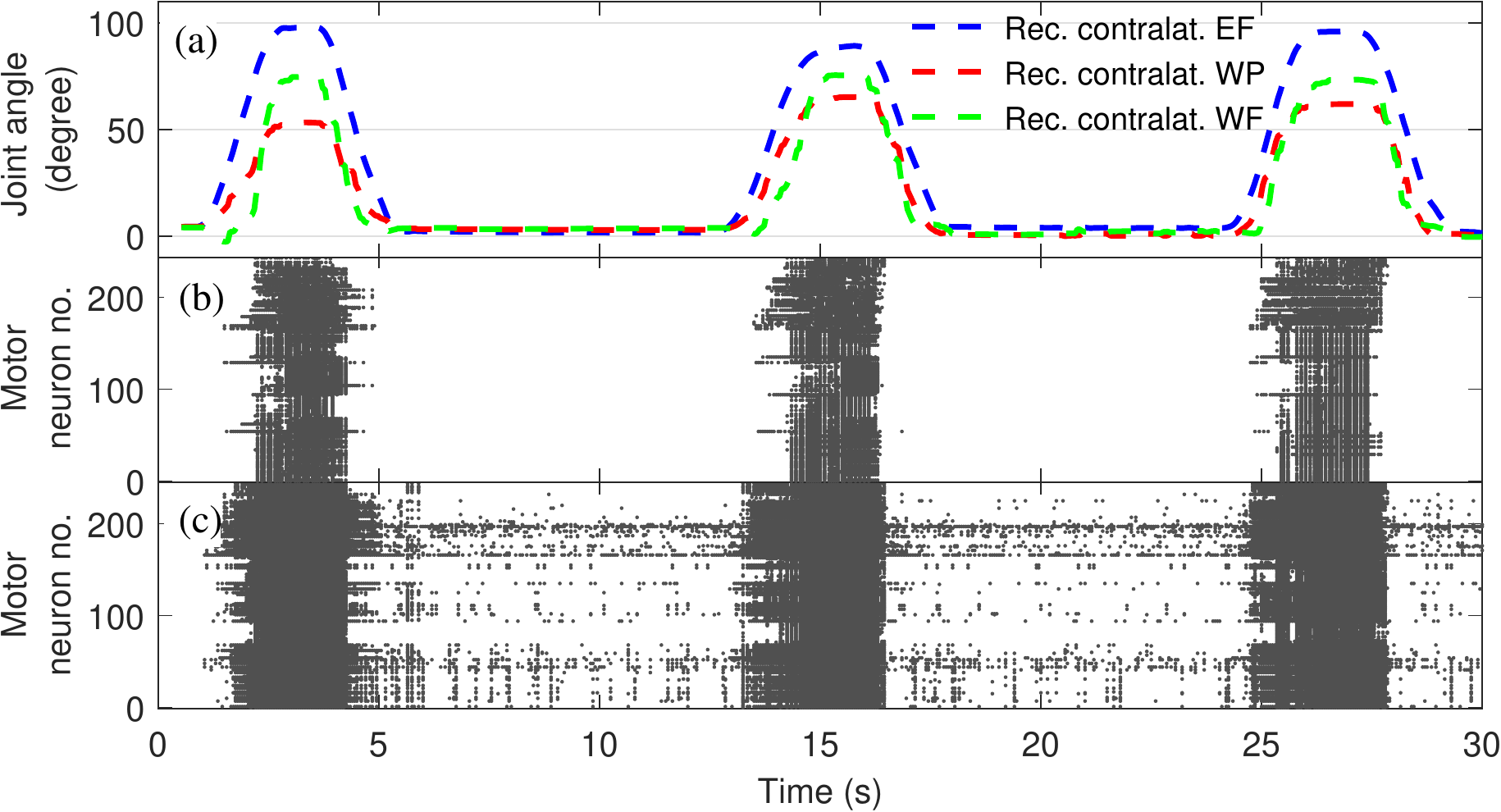}
\caption{(a) Recorded contralateral 3-DoF movements (case 4) and corresponding raster plots of (b) extended motor neuron spike trains extracted by 2-class K-means algorithm versus (c) that extracted by the adaptive threshold.}
         \label{fig:10}
\end{figure}

As discussed earlier we used an adaptive thresholding procedure to extract motor neuron discharge timings and decoded the new set of MUSTs to estimate the kinematics. Here, we show how much improvement on EMG decomposition and movement estimation is achieved using the new method, by comparing its performance against fixed IPT thresholding given by the 2-class K-means algorithm. Figure \ref{fig:10} compares the raster plots of the extracted MUST channels when using fixed and adaptive thresholds. It also shows the corresponding desired movements. Obviously, the new method avoided many missed detections. The adaptive IPT threshold could even extract discharge timings with lower amplitude when action potentials change in shapes in dynamic conditions.

\begin{figure}
\centering
\includegraphics[scale=0.65]{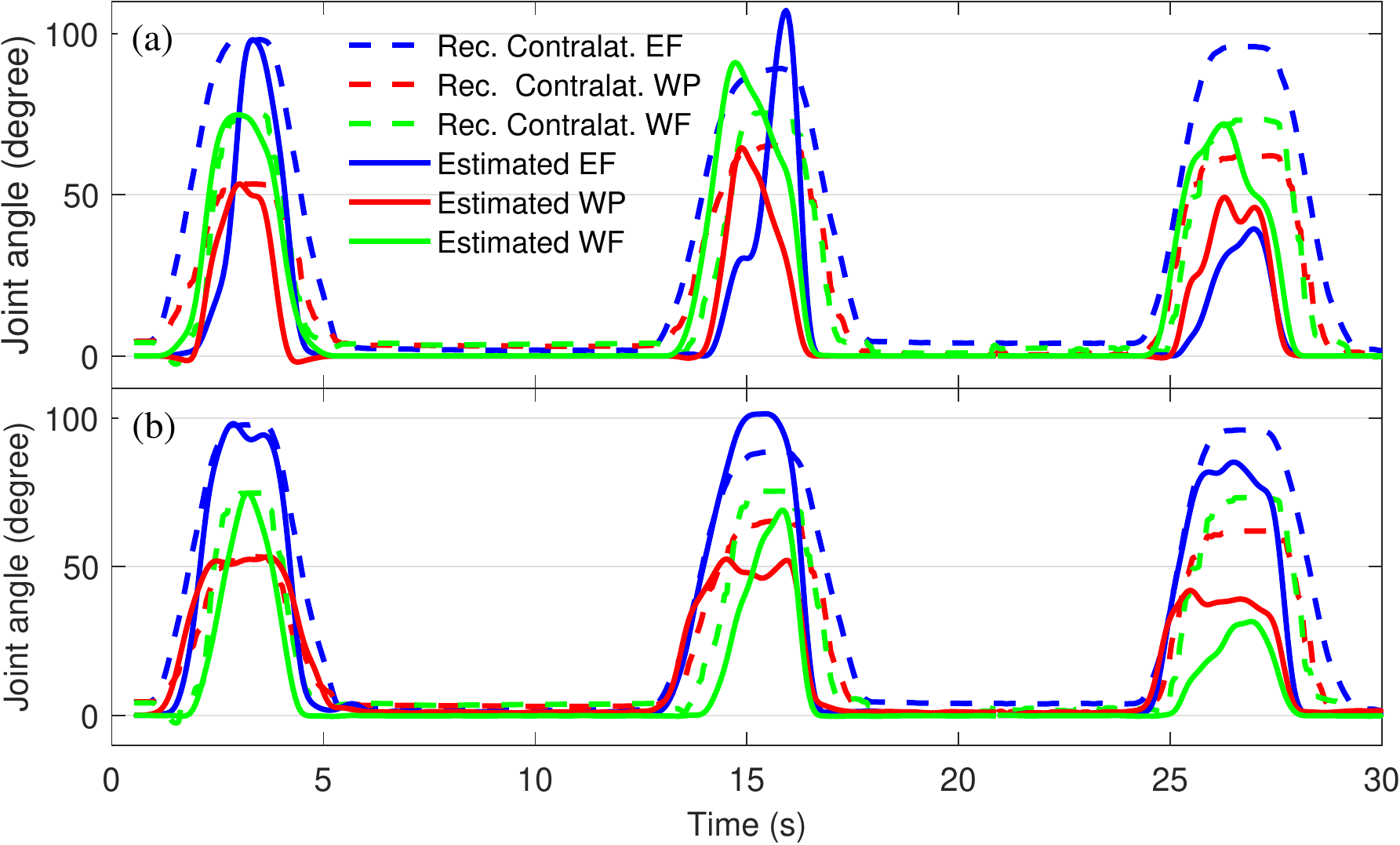}
\caption{Kinematic Estimation for a 3-DoF movement using the set of MUSTs extracted by (a) 2-class K-means algorithm versus (b) the adaptive threshold algorithm.}
         \label{fig:11}
\end{figure}

Figure \ref{fig:11} compares the kinematic estimation using the two set of MUSTs, which reveals the new method improves movement estimation. The reported $R^2$ values indicate 28\% of improvement in projection on both the training and test sets.

\section{Discussion}
We proposed a new framework for prosthetic hand control based on decoding of discharge timings of spinal motor neurons for intuitive estimation of intended kinematics. The high dimensional data of motor neuron discharge timings, extracted by EMG decomposition, was projected by a linear unsupervised metric learning approach to the lower dimensional space of active DoFs. The use of a dimensionality reduction algorithm is justified by the way the neuro-muscular system works, where dozens of motor neurons excite thousands of muscle fibers in each muscle, to control only two joints in that muscle for a single contraction.

We chose metric learning deliberately, as it complies with muscle synergies for natural control and movement. We know there is a one-to-one relationship between a muscle and a motor pool. However, for most voluntary movements, a few muscles are activated and coordinated together to perform the task. This means that the whole mapping from motor neurons to movements is not one-to-one. Metric learning can independently approximate weighting vectors for each degree of freedom, and those weightings do not force a zero-or-one contribution of input motor units in different axes of the projection space. Hence, it enables an individual input to contribute in control of more than one DoF. The loading vectors in figure \ref{fig:6} shows that some input channels of motor unit discharge timings have non-zero weights in even two or three DoFs.

The linear algorithm employed here is consistent with both Henneman's size principle \citep{henneman1965} and the rate coding \citep{Conwit1999} that govern the relationship between motor neuron activity and the force generated by muscle contraction. A motor neuron pool comprises neurons with different thresholds of excitability, distributed across the full range of force. When these motor neurons receives a common input, they generate a neural drive to the muscle which is the ensemble of discharge timings of the active motor neurons and which is linearly associated to force (force is the low-pass filtered version of the neural drive to muscle). This fact justifies that a linear combination of input channels that correspond to different types of motor units is able to mimic the natural solution that provides smooth, precise, and consistent modulation of contraction levels from just increased or decreased levels of synaptic inputs.

The data used in this work include movements with up to three DoFs, and we used the first trial of movements to detect the active DoFs. PCA has the potential to control many degrees of freedom. However, a major improvement is to find an automatic way to assign the rotated principal components to an active DoF. For this purpose, we propose to use the sparsified loadings (see figure \ref{fig:6}), as we noticed they provide important information about active DoFs. From these rotated loading it is easy to find decomposed sources that only fire during a single type of movement. Once we know the active DoFs, the order can be learned very quickly by a query-based algorithm to find the true permutation, a problem very similar to the famous mastermind game \citep{knuth1976, chvatal1983}, with much lower complexity as the active DoFs are already known from the discriminative MUs given by the loadings.

In the proposed projection algorithm, each individual principal component in the subspace will represent just one direction of movement. However, to perfectly follow the conceptual projection operator given in figure \ref{fig:1}, we need to modify the algorithm such that movements created by antagonistic pairs of muscles appear in the same eigen-direction. We have previously tried this idea by using PCA decomposition in the complex domain of MUSTs to capture the motor synergism \citep{andalib2016}. We showed complex PCA can project our reconstructed complex MUSTs of a simple wrist flexion-extension to only one direction in the LDS.

Because of this intuitive control strategy, we believe the proposed decoding approach can be naturally used in a more general framework based on human perception-action cycle (PAC) for motor control, which means it can be used in a closed-loop control setup of a bidirectional prosthesis that can also feedback environmental, say tactile, information to the PNS. Our computationally simple decoding algorithm facilitates online real-time implementations as required for the PAC. 

\ack
This work is sponsored in part by the Defense Advanced Research Projects Agency (DARPA) Electrical Prescriptions (ElectRx) program under the auspices of Dr.~Eric~Van~Gieson through the Space and Naval Warfare Systems Center, Pacific Grant/Contract No.~N66001-15-C-4018 to the University of Florida. This text is approved for Public Release by DARPA, Distribution Unlimited.

This work is supported in part by the European Union's Horizon 2020 research and innovation programme under the Marie Skłodowska-Curie grant agreement No.~702491 (NeuralCon, FN).

\newcommand{\newblock}{}
\bibliographystyle{dcu}

\bibliography{references}

\end{document}